\documentclass[sn-mathphys-ay]{sn-jnl}

\usepackage{graphicx}%
\usepackage{multirow}%
\usepackage{amsmath,amssymb,amsfonts}%
\usepackage{amsthm}%
\usepackage{mathrsfs}%
\usepackage[title]{appendix}%
\usepackage{xcolor}%
\usepackage{textcomp}%
\usepackage{manyfoot}%
\usepackage{booktabs}%
\usepackage{algorithm}%
\usepackage{algorithmicx}%
\usepackage{algpseudocode}%
\usepackage{listings}%
\usepackage{lineno}
\usepackage[colorinlistoftodos]{todonotes}

\usepackage{graphicx, caption, subcaption}

\newcommand{\EF}{{\mathrm{eff}}}
\newcommand{\FF}{{\mathrm{ff}}}
\newcommand{\PM}{{\mathrm{pm}}}
\newcommand{\TR}{{\mathrm{tr}}}

\newcommand{\BL}{{\mathrm{bl}}}

\newcommand{\WALL}{{\mathrm{wall}}}

\newcommand{\OUTFLOW}{{\mathrm{outflow}}}

\renewcommand{\vec}[1]{{\ensuremath{\boldsymbol{\mathrm #1}}}}
\newcommand{\ten}[1]{\ensuremath{\boldsymbol{\mathsf{#1}}}}

\theoremstyle{thmstyleone}%
\theoremstyle{thmstyletwo}%
\newtheorem{remark}{Remark}%

\theoremstyle{thmstylethree}%

\begin{document}

\title[A hybrid-dimensional Stokes--Brinkman--Darcy model]{A hybrid-dimensional Stokes--Brinkman--Darcy model for arbitrary flows to the fluid--porous interface}


\author[1]{\fnm{Linheng} \sur{Ruan}}\email{linheng.ruan@ians.uni-suttgart.de}

\author*[1]{\fnm{Iryna} \sur{Rybak}}\email{iryna.rybak@ians.uni-stuttgart.de}

\affil[1]{\orgdiv{Institute of Applied Analysis and Numerical Simulation},
\orgname{University of Stuttgart}, \orgaddress{\street{Pfaffenwaldring 57}, \city{Stuttgart}, \postcode{70569}, \country{Germany}}}


\abstract{
Mathematical modelling of coupled flow systems containing a free-flow region in contact with a porous medium is challenging, especially for arbitrary flow directions to the fluid--porous interface. Transport processes in the free flow and porous medium are typically described by distinct equations: the Stokes equations and Darcy's law, respectively, with an appropriate set of coupling conditions at the common interface. Classical interface conditions based on the Beavers--Joseph condition are not accurate for general flows. Several generalisations are recently developed for arbitrary flows at the interface, some of them are however only theoretically formulated and still need to be validated. 

In this manuscript, we propose an alternative to couple free flow and porous-medium flow, namely, the hybrid-dimensional Stokes--Brinkman--Darcy model. Such formulation incorporates the averaged Brinkman equations within a complex interface between the free-flow and porous-medium regions. The complex interface acts as a buffer zone facilitating storage and transport of mass and momentum and the model is applicable for arbitrary flow directions. We validate the proposed hybrid-dimensional model against the pore-scale resolved model in multiple examples and compare numerical simulation results also with the classical and generalised coupling conditions from the literature. The proposed hybrid-dimensional model demonstrates its applicability to describe arbitrary coupled flows and shows its advantages in comparison to other generalised coupling conditions. 
}



\maketitle

\subsection*{Article Highlights}

\begin{itemize}
\item New hybrid-dimensional model is presented to couple the Stokes equations and Darcy's law.
\item The proposed model is validated numerically using pore-scale resolved simulations and compared to other coupling concepts.
\item Higher-order correction terms in the hybrid-dimensional model enhance the accuracy and make it suitable for arbitrary flows.
\end{itemize}

\section{Introduction}\label{sec1}
Study of flow systems involving free flow and porous medium is an active research area with applications in biology, agriculture, and industry, such as drug transport in biological tissues, subsurface drainage, and industrial filtration. Fluid behaviour varies significantly in these regions, and it can be described from either pore-scale or macroscale perspective. For the pore-scale modelling, the pore geometry is resolved and the Stokes equations are considered in the free-flow domain and the pore space of the porous medium~\citep{Hornung_97, Jaeger_Mikelic_09, weishaupt2019efficient, Lacis_etal_20, Eggenweiler_Rybak_20}. However, this approach is computationally extensive for applications, and it is mainly used for model validation and calibration~\citep{Rybak_etal_19}.

From the macroscale perspective, laminar flow in the plain--fluid region is described using the Stokes equations, while flow in the porous medium is governed by Darcy's law. Modelling the interaction between these two flow subdomains is a challenging problem. Extensive studies have been conducted on coupling the Stokes and Darcy's models, which can be modelled by considering either a sharp interface or a transition zone between the free flow and porous medium~\citep{OCHOATAPIA19952635, Goyeau_Lhuillier_etal_03, Angot_etal_17}.

In the last decades, most of the studies have been focused on coupling conditions at the sharp fluid--porous interface based on the Beavers--Joseph condition~\citep{Beavers_Joseph_67, Saffman, Jaeger_Mikelic_00, Jaeger_Mikelic_09,Discacciati_Miglio_Quarteroni_02, Discacciati_Quarteroni_09, Layton_Schieweck_Yotov_03, Girault-Riviere-09}. However, these conditions are limited to the cases where the flow is either parallel or perpendicular to the interface~\citep{Eggenweiler_Rybak_20}. To address these limitations, several generalised interface conditions have been introduced using different methods, e.g., asymptotic modelling, homogenisation, boundary layer theory and upscaling techniques~\citep{OCHOATAPIA19952635, Jaeger_etal_01, ValdesParada_etal_09, Carraro_etal_15,  Angot_etal_17, Lacis_Bagheri_17, angot2018well, Lacis_etal_20, Angot_etal_20, Eggenweiler_Rybak_MMS20, Strohbeck-Eggenweiler-Rybak-23}. In these generalised conditions, additional terms are often introduced, but some of the conditions are only theoretically derived without accompanying any numerical simulation results. Recently, generalised interface conditions have been developed using homogenisation and boundary layer theory~\citep{Eggenweiler_Rybak_MMS20}. This approach involves some additional terms, making it suitable for flows in any direction compared to the classical Beavers--Joseph condition.

As an alternative approach, we can also consider a transition zone between the free-flow and porous-medium subdomains. Compared to the sharp interface, the transition zone stores  mass, momentum, and energy, and facilitates their transport in the tangential direction. The thickness of the transition zone is always considered to be narrow compared to the entire domain size \citep{Angot_etal_17, Ruan-Rybak-23}. Therefore, it can be regarded as a lower-dimensional inclusion and treated as a complex interface. Such hybrid-dimensional modelling is widely used in fractured porous-medium systems \citep{Lesinigo_etal_11, brenner-etal-2018, Rybak-Metzger-20, gander-hennicker-masson-21, gander2023dimensional,DUGSTAD2022104140, maxi2024}, and it is proven to be an effective approach in comparison to the full-dimensional formulation. In our previous work, we derived the hybrid-dimensional Stokes--Brinkman--Darcy model, which consists of the averaged Brinkman equations in the transition zone and the corresponding transmission conditions~\citep{Ruan-Rybak-23, Ruan-Rybak-2024}. The new model allows the treatment of viscous flows along the complex interface, and involves additional higher-order correction terms. We proved the well-posedness of this hybrid-dimensional model and provided the numerical convergence study using analytical solutions~\citep{Ruan-Rybak-2024}.

In this paper, we aim to provide thorough numerical study of the developed hybrid-dimensional model in order to validate its applicability for arbitrary flow directions and to make inter-comparison with the classical and generalised interface conditions. Various scenarios are tested to demonstrate the suitability of the developed model for non-parallel flows in the vicinity of the fluid--porous interface.

The paper is arranged as follows: In section~\ref{sec2}, we introduce the pore-scale model and the macroscale models for the coupled free-flow and porous-medium system with three different coupling concepts, namely, the hybrid-dimensional model, the classical interface conditions and the generalised interface conditions. Numerical simulation results for three different test cases (validation against analytical solutions, lid-driven cavity over a porous bed and splitting flow) are provided in section~\ref{sec3:NumSimRes}. Finally, discussion and conclusions follow in  section~\ref{sec:discussion}.

\section{Coupled flow models}\label{sec2}
\subsection{Flow domain and geometry}\label{sec2-1}
\begin{figure}[h]
\centering
\includegraphics[width=0.68\textwidth]{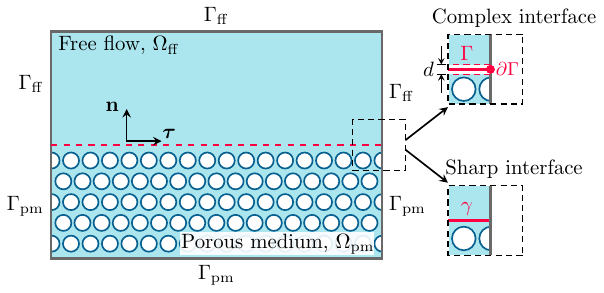}
\caption{Geometry of the coupled free-flow and porous-medium system}\label{fig1}
\end{figure}

The coupled flow domain $\Omega= \Omega_\FF \cup \Omega_\PM\subset \mathbb{R}^n$ consists of two subdomains: the free flow $\Omega_\FF$ and the porous medium $\Omega_\PM$ (Fig.~\ref{fig1}). A steady-state flow of a single-phase and incompressible fluid with constant viscosity is taken into account in this work. The same fluid fully saturates the porous-medium subdomain, where the solid inclusions are periodically distributed. Note that periodicity is not requested for the proposed hybrid-dimensional model, but is taken here to be able to compare the model with the generalised interface conditions in section~\ref{sec:generalIC}. The Reynolds number is assumed to be low such that the inertial effects are neglected.

In the pore-scale setting, there is no transition region or interface between the two subdomains (Fig.~\ref{fig1}, left). The pore geometry is resolved and we denote the pore space in the porous medium as $\Omega_{\PM,\epsilon}\subset\Omega_\PM$. The Stokes equations are considered in the fluid domain $\Omega_\epsilon := \Omega_\FF \cup \Omega_{\PM,\epsilon}$. From the macroscale perspective, the two subdomains are considered as two distinct continua. Either a sharp interface or a transition zone is considered between the free flow and the porous medium. In the former case, the macroscale model is coupled at the sharp fluid--porous interface $\gamma = \partial \Omega_\FF \cap \partial \Omega_\PM$ (Fig~\ref{fig1}). A local coordinate system is defined with the unit tangential vector $\vec{\tau}$ and unit normal vector $\vec{n}$. In the latter case, the free-flow and porous-medium subdomains ($\Omega_\FF \cap \Omega_\PM  = \emptyset$) are separated by a transition zone $\Gamma= \left\{\vec{x}\in \mathbb{R}^2 \big| \vec{x}= \vec{s} \pm 1/2 t d (\vec{s}) \vec{n}, t \in [0,1], \vec{s}\in \gamma\right\}$. We assume that the thickness $d>0$ is sufficiently small so that we can treat the transition zone as a complex interface $\Gamma$ (Fig.~\ref{fig1}) in the hybrid-dimensional setting.

\subsection{Pore-scale model}
In the pore-scale model, the flow in the fluid domain $\Omega_\epsilon$ is governed by the Stokes equations 
\begin{align}
    \nabla \cdot \vec{v}_\epsilon = 0 \hspace*{+1ex} \quad&\textnormal{in } \Omega_\epsilon, \label{eqn:porescaleStokes1}\\
    -\nabla \cdot \ten{T}\left(\vec{v}_\epsilon, p_\epsilon \right) = \vec{f}_\epsilon \quad &\textnormal{in }\Omega_\epsilon,\label{eqn:porescakeStokes2}\\
    \vec{v}_\epsilon = 0 \hspace*{+1ex} \quad&\textnormal{on }  \partial \Omega_\epsilon / \partial \Omega,\label{equ:fluid-solidBC}
\end{align}
where $\vec{v}_\epsilon$ and $ p_\epsilon$ denote the pore-scale velocity and pressure, $\vec{f}_\epsilon$ is the source term and $\ten{T}\left(\vec{v},p\right):= \mu \nabla \vec{v}- p \ten{I}$ represents the stress tensor with the dynamic viscosity $\mu>0$. In Eq.~\eqref{equ:fluid-solidBC}, we prescribe the non-slip boundary condition at the fluid--solid interface. On the external boundary $\partial \Omega  = \partial \Omega_D \cup \partial \Omega_N$, $\partial \Omega_D \cap \partial\Omega_N\ne \emptyset$, we consider either Dirichlet (D) or Neumann (N) boundary conditions
\begin{equation}
    \vec{v}_\epsilon =  \overline{\vec{v}}\textnormal{ on } \partial \Omega_D,\quad \ten{T}\left(\vec{v}_\epsilon ,p_\epsilon\right) \cdot \vec{n}_{\Omega} =\overline{\vec{t}} \textnormal{ on } \partial \Omega_N, \label{eqn:porescaleBC}
\end{equation}
with the given values $\overline{\vec{v}}$ and $\overline{\vec{t}}$, and the unit outward normal vector $\vec{n}_\Omega$ on $\partial \Omega$. We use the pore-scale model for validation of macroscale coupled models.

\begin{figure}[h]
\centering
\includegraphics[width=0.68\textwidth]{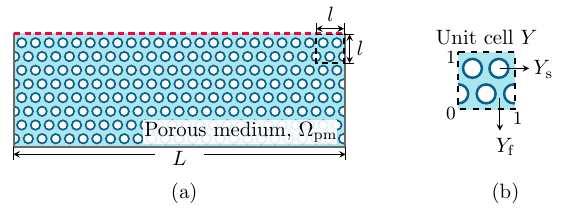}
\caption{Porous-medium geometry with periodic solid inclusions (a) and unit cell (b)}\label{fig2}
\end{figure}

In the porous-medium region, we apply homogenisation theory \citep{Hornung_97} to calculate the effective permeability values for the macroscale model. The pore-scale and macroscale are assumed to be separable that means the scale separation parameter $\epsilon := l/ L \ll 1$ for the characteristic pore size $l$ and the length of the domain $L$ (Fig.~\ref{fig2}a). The porous-medium region is constructed by the repetition of the unit cell $Y$ scaled by $\epsilon$. The unit cell consists of the fluid part $Y_\mathrm{f}$ and the solid part $Y_\mathrm{s}$ (Fig.~\ref{fig2}b). Flow behaviour in the porous medium follows Darcy's law when $\epsilon\rightarrow0$.

The permeability tensor $\ten{K}$ in the unit cell is given by
\begin{equation}
    \ten{K} : = \left(k_{ ij}\right)_{1\leq i,j \leq n}= \int_{Y_\mathrm{f}} w_{ji}~\mathrm{d}\vec{y}, \quad \vec{y}:=\frac{\vec{x}}{\epsilon}\in Y, \label{eqn:unitcellpermeability}
\end{equation}
where $\vec{w}_j:= \left(w_{j1}, \dots, w_{jn}\right)^\top$ is the solution of the following cell problem
\begin{align}
    -\Delta_{\vec{y}} \vec{w}_j + \nabla_{\vec{y}} \pi_j= \vec{e}_j \quad&\textnormal{in } Y_\textrm{f}, \label{eqn:cellproblem1}\\
    \nabla_\vec{y} \cdot \vec{w}_j= 0 \quad &\textnormal{in } Y_\mathrm{f},\label{eqn:cellproblem2}\\ 
    \vec{w}_j = \vec{0}  \quad&\textnormal{on } \partial Y_\mathrm{f} / \partial Y.\label{eqn:cellproblem3}
\end{align}
The functions $\{\vec{w}_j, \pi_j\}$ are $1$-periodic in  $\vec{y}$ and $ \int_{Y_\mathrm{f}} \pi_j~\mathrm{d}\vec{y}=0$, where $\pi_j$, $j\in\{1,\dots,n\}$ is the pressure in the unit cell.
The physical permeability tensor is obtained by scaling $\ten{K}$ as follows $\ten{K}_{\PM}:= \epsilon^2 \ten{K}$.

\subsection{Macroscale models in two flow domains}\label{sec:macroscale}
In this section, we present the macroscale models for the free flow and porous medium, where the macroscale velocity and pressure are denoted by $\left(\vec{v}_i, p_i\right)$ for $i\in\{\FF, \PM\}$. Under the assumptions introduced in section~\ref{sec2-1}, the Stokes equations are used to describe the laminar flow in the free-flow subdomain
\begin{align}
    \nabla \cdot \vec{v}_\FF &= 0 \hspace{+3.5ex} \textnormal{ in } \Omega_\FF, \label{equ:Stokes1}\\ 
    -\nabla \cdot \ten{T} (\vec{v}_\FF, p_\FF) &= \vec{f}_\FF \quad \textnormal{ in } \Omega_\FF, \label{equ:Stokes2}
\end{align}
where $\vec{f}_\FF$ is the source term. On the external boundary $\Gamma_\FF:=\Gamma_{\FF, D} \cup\Gamma_{\FF,N}$, where $\Gamma_{\FF, D}\cap\Gamma_{\FF, N} \ne \emptyset$ and $\Gamma_{\FF,D} \ne \emptyset$, the following boundary conditions are considered
\begin{equation}
    \vec{v}_\FF = \overline{\vec{v}}_\FF \textnormal{ on } \Gamma_{\FF, D}, \quad \ten{T}\left(\vec{v}_\FF, p_\FF\right) \cdot \vec{n}_\FF = \overline{\vec{t}}_\FF \textnormal{ on }\Gamma_{\FF, N}, \label{eqn:FFBC}
\end{equation}
with the unit outward vector $\vec{n}_\FF$ and the given boundary data $\overline{\vec{v}}_\FF$, $\overline{\vec{t}}_\FF$. 

In the porous-medium subdomain, the slow flow is governed by Darcy's law 
\begin{align}
    \nabla \cdot \vec{v}_\PM &= f_\PM \hspace{+11ex} \textnormal{ in } \Omega_\PM,\label{equ:Darcy1}\\ 
    \vec{v}_\PM &= -\frac{\ten{K}_\PM}{\mu} \nabla p_\PM \quad \textnormal{ in } \Omega_\PM, \label{equ:Darcy2}
\end{align}
with the porous-medium source term $f_\PM$. The following boundary conditions
\begin{equation}
     p_\PM = \overline{p}_\PM \quad \textnormal{on } \Gamma_{\PM, D}, \quad \vec{v}_\PM \cdot \vec{n}_\PM = \overline{v}_{\PM,n}\quad \textnormal{on } \Gamma_{\PM,N} ,\label{equ:PMBC}
\end{equation}
are taken into account on the external boundary $\Gamma_\PM=\Gamma_{\PM,D} \cup \Gamma_{\PM,N}$. Here, $\overline{p}_\PM$ and $\overline{v}_{\PM,n}$ are the given functions and $\vec{n}_\PM$ denotes the unit outward vector on $\Gamma_\PM$. 

\subsubsection{Hybrid-dimensional Stokes--Brinkman--Darcy model}\label{sec:hybrid}
In this section, we present the hybrid-dimensional model derived in our previous work~\citep{Ruan-Rybak-23,Ruan-Rybak-2024}, where a thin transition zone is considered between the two subdomains. 
The model contains the averaged Brinkman equations for the transition region, which is modelled as a complex interface of co-dimension one, and the corresponding transmission conditions. 


On the complex interface $\Gamma$, the averaged Brinkman equations are given by
\begin{align}
    \vec{v}_\FF \cdot  \vec{n} |_{\gamma_\FF} - \vec{v}_\PM \cdot \vec{n} |_{\gamma_\PM}  &= - d \frac{\partial  V_{\vec{\tau}} }{\partial \vec{\tau}} \hspace{+35.5ex}\textnormal{ on } \Gamma, \label{eqn:AverageBrinkmanMass}\\    
    \Big(  \vec{n} \cdot  \ten{T}\left(\vec{v}_\FF, p_\FF \right) \cdot  \vec{n} -\frac{\mu }{\sqrt{K_{\Gamma}}}& \left(\vec{\beta}\vec{v}_\FF \right)\cdot\vec{n} \Big)\Big|_{\gamma_\FF} + p_\PM \big|_{\gamma_\PM } \nonumber\\
    &= d \Big(  
    \mu (\ten{K}_{\Gamma}^{-1} \vec{V})\cdot \vec{n}- \mu_\EF \frac{\partial^2 V_{\vec{n}}}{\partial \vec{\tau}^2}-F_{\vec{n}}\Big) \hspace{+9ex}\textnormal{ on } \Gamma,\label{eqn:AveragedBrinkmanMomentumNormal} \\
    \Big(\vec{n} \cdot \ten{T}\left(\vec{v}_\FF, p_\FF \right) \cdot  \vec{\tau} -  \frac{\mu }{\sqrt{K_{\Gamma}}} &\left(\vec{\beta}\vec{v}_\FF \right) \cdot  \vec{\tau} \Big)\Big|_{\gamma_\FF}-
        \frac{\alpha \mu_\EF (6 V_{\vec{\tau}}- 2 \vec{v}_\FF \cdot  \vec{\tau}|_{\gamma_\FF})}{\alpha d +4\sqrt{K_\PM}}\nonumber \\
        &= d \Big( 
        \mu (\ten{K}_{\Gamma}^{-1} \vec{V}) \cdot  \vec{\tau}
        -  \mu_\EF \frac{\partial^2 V_{\vec{\tau}}}{\partial \vec{\tau}^2} + \frac{\partial P}{\partial \vec{\tau}}-F_{\vec{\tau}} \Big) \quad \textnormal{ on } \Gamma,\label{eqn:AveragedBrinkmanMomentumTangential}
\end{align}
where $\mu_\EF>0$ is the effective viscosity, $\alpha>0$ is the slip coefficient, and $\vec{\beta}$ is the stress jump tensor. The averaged velocity and pressure are defined as $\vec{V}:= \frac{1}{d}\int^{d/2}_{-d/2} \vec{v}_{\Gamma}~\mathrm{d}n$ and $P:=\frac{1}{d}\int^{d/2}_{-d/2} p_{\Gamma}~\mathrm{d}n$ using velocity and pressure $(\vec{v}_{\Gamma},p_{\Gamma})$ in the transition region. 
Interfaces $\gamma_\FF$ and $\gamma_\PM$ are defined at the top and bottom of the transition region in the full-dimensional model~\citep{Ruan-Rybak-2024}, and they are fictitious in the hybrid-dimensional model.  
The normal and tangential components of the averaged velocity are denoted by $V_\vec{n}:=\vec{V}\cdot\vec{n}$ and $V_\vec{\tau}: = \vec{V}\cdot \vec{\tau}$, and the corresponding source terms are $F_\vec{n}$, $F_\vec{\tau}$. The permeability tensor $\ten{K}_\Gamma$ is symmetric positive definite,  $K_\Gamma:=\|\ten{K}_\Gamma\|_{\infty}$, and $K_\PM:= \vec{\tau} \cdot \ten{K}_\PM \cdot \vec{\tau}$. 

The effective viscosity $\mu_\EF$ in \eqref{eqn:AveragedBrinkmanMomentumNormal}--\eqref{eqn:AveragedBrinkmanMomentumTangential} comes from the Brinkman equations, and it differs from intrinsic viscosity due to the dispersion of viscous diffusion flux. It has been generally accepted that $\mu_\EF$ is dependent on the type of porous media as well as the strength of the flow. Through volume averaging of the Navier--Stokes equations~\citep{OCHOATAPIA19952635}, the effective viscosity is set to $\mu_\EF=\mu/\phi$ with porosity $\phi$.  The permeability values in the transition zone $\ten{K}_\Gamma$ are expected to be larger or equal to the porous-medium permeability $\ten{K}_\PM$. The stress jump parameter $\ten{\beta}$ is assumed to be positive semi-definite and need to be determined. The slip coefficient $\alpha$ comes from the Beavers--Joseph--Saffman condition, which is typically considered $\alpha=1$ in the literature, e.g.~\citep{Layton_Schieweck_Yotov_03, Discacciati_Quarteroni_09}.

On the external boundary $\partial \gamma = \partial \gamma_D \cap \partial \gamma_N$, we consider the following boundary conditions
\begin{equation}
    \vec{V} = \overline{\vec{V}} \hspace*{+1ex}\textnormal{ on }\partial \gamma_D , \quad    \mu_\EF \frac{\partial V_\vec{n}}{\partial \vec \tau}= \overline{T}_{\vec{n}}, \quad \mu_\EF \frac{\partial V_\vec{\tau}}{\partial \vec \tau}-P=\overline{T}_\vec{\tau} \hspace*{1ex}\textnormal{ on } \partial \gamma_N, \label{eqn:gammaBC}
\end{equation}
where $\overline{\vec{V}}$, $\overline{T}_\vec{n}$ and $\overline{T}_\vec{\tau}$ are the given data. 

\begin{remark}
    Note that the averaged Brinkman equations~\eqref{eqn:AverageBrinkmanMass}--\eqref{eqn:AveragedBrinkmanMomentumTangential} are the second order partial differential equations of dimension $(n-1)$. If there is no transition zone $(d=0)$, we recover from Eq.~\eqref{eqn:AverageBrinkmanMass} the mass conservation equation~\eqref{eqn:BJ1} and from Eq.~\eqref{eqn:AveragedBrinkmanMomentumNormal} the normal stress jump condition between the free flow and porous medium \citep{OCHOATAPIA19952635,  Angot_etal_17}, respectively. 
\end{remark}

In the hybrid-dimensional problem, the averaged Brinkman equations on $\Gamma$ are not able to extrapolate the velocity and the pressure values on the top $\gamma_\FF$ and bottom $\gamma_\PM$ of the transition zone. Therefore, to obtain a closed model, we derive the following transmission conditions
\begin{align}
     \Big( \vec{n} \cdot  \ten{T}(\vec{v}_\FF&, p_\FF ) \cdot  \vec{n}-\frac{\mu }{\sqrt{K_{\Gamma}}} \left(\vec{\beta}\vec{v}_\FF \right)\cdot\vec{n} \Big)\Big|_{\gamma_\FF}\nonumber\\
    &= -\frac{\mu_\EF}{d}\left((\lambda_1+\lambda_2) V_{\vec{n}}  - \lambda_1 \vec{v}_\FF\cdot \vec{n} |_{\gamma_\FF}
      - \lambda_2 \vec{v}_\PM \cdot \vec{n} |_{\gamma_\PM}\right)-P,\label{eqn:TransmissionGammaFFNormal}
      \\
    \Big(\vec{n}\cdot \ten{T}(\vec{v}_\FF&, p_\FF ) \cdot  \vec{\tau}-\frac{\mu }{\sqrt{K_{\Gamma}}} \left(\vec{\beta}\vec{v}_\FF \right) \cdot  \vec{\tau} \Big)\Big|_{\gamma_\FF}\nonumber\\
    &= -\frac{\mu_\EF}{d}\left( \frac{ 6( \alpha d + 2\sqrt{K_\PM}) }{\alpha d+4\sqrt{K_\PM}} V_{\vec{\tau}} - \frac{  4 ( \alpha d + 3\sqrt{K_\PM}) }{ \alpha d+4\sqrt{K_\PM} } \vec{v}_\FF \cdot  \vec{\tau}  \big|_{\gamma_\FF}\right),\label{eqn:TransmissionGammaFFtangential}\\
    p_\PM \big|_{\gamma_\PM } &= -\frac{\mu_\EF}{d} \left((\lambda_1+\lambda_2)V_{\vec{n}}-\lambda_2\vec{v}_\FF \cdot \vec{n}|_{\gamma_\FF} -\lambda_1\vec{v}_\PM \cdot \vec{n} |_{\gamma_\PM} \right) + P,\label{eqn:TransmissionGammaPM}
\end{align}
using the \textit{a priori} assumptions on the velocity and pressure profiles across the transition region. We assume a quadratic tangential velocity $\vec{v}_\Gamma \cdot \vec{\tau}$ and a constant pressure $p_{\Gamma}$ profiles while deriving the transmission conditions~\eqref{eqn:TransmissionGammaFFNormal}--\eqref{eqn:TransmissionGammaPM}. The dimensionless numbers $\lambda_1$ and $\lambda_2$ in Eq.~\eqref{eqn:TransmissionGammaFFNormal}, \eqref{eqn:TransmissionGammaPM} determine the profile of the normal velocity. In this work, we consider the linear $(\lambda_1 = 2,\, \lambda_2 = 0)$, piecewise linear $(\lambda_1 = 3,\, \lambda_2 = 1)$ and quadratic $(\lambda_1 = 4,\, \lambda_2 = 2)$ normal velocity profiles.

\begin{remark}
    According to \citet{Ruan-Rybak-2024}, the parameter $\ten{\beta}$ characterizes the surface roughness at the boundary between the free flow and the complex interface, while the coefficient $\alpha$ represents the roughness between the complex interface and the porous medium. The complex interface acts as a buffer zone, where arbitrary velocities from the free-flow region are decelerated and enter the porous medium subdomain. This buffering mechanism facilitates a smooth and accurate exchange of mass and momentum between the regions, enhancing the model ability to handle diverse flow conditions. 
\end{remark}

\subsubsection{Classical interface conditions}\label{sec:classicalIC}
In this section, we introduce the classical coupling conditions typically used in the literature for coupling the Stokes and Darcy equations~\citep{Discacciati_Miglio_Quarteroni_02, Layton_Schieweck_Yotov_03, Riviere}. These conditions are composed of the mass conservation, the balance of normal  forces across the interface and the Beavers--Joseph condition~\citep{Beavers_Joseph_67}: \begin{align}
\vec{v}_\FF \cdot \vec{n}&= \vec{v}_\PM \cdot \vec{n} \hspace{+10.5ex} \textnormal{ on }\gamma,\label{eqn:BJ1}\\
-\vec{n} \cdot \ten{T} (\vec{v}_\FF, p_\FF) \cdot \vec{n } &= p_\PM \hspace{+13.8ex} \textnormal{ on }\gamma,\label{eqn:BJ2}\\
\vec{v}_\FF \cdot \vec{\tau} - \vec{v}_\PM \cdot \vec{\tau} &= \frac{\sqrt{K_\PM}}{\alpha_\mathrm{BJ}} \frac{\partial \vec{v}_\FF}{\partial \vec{n}} \cdot\vec{\tau} \quad \textnormal{ on }\gamma,\label{eqn:BJcondition}
\end{align}
where $\alpha_{BJ}>0$ is the Beavers--Joseph coefficient. Note that these conditions are developed for parallel flows. 

\subsubsection{Generalised interface conditions}\label{sec:generalIC}
In this section, we provide the generalised interface conditions derived using homogenisation and boundary layer theory~\citep{Eggenweiler_Rybak_MMS20}, which are applicable for arbitrary flows near the sharp interface. These generalised conditions are given by
\begin{align}
    \vec{v}_\FF \cdot \vec n     &= \vec{v}_\PM \cdot \vec{n}  \quad &\textnormal{on } \gamma, \label{equ:ER1}\\
    - \vec{n} \cdot \ten{T}(\vec{v}_\FF, p_\FF) \cdot \vec{n}+ \mu N_\mathrm{s}^\BL \frac{\partial \vec{v}_\FF}{\partial \vec{n}} \cdot\vec{\tau} &=p_\PM \quad& \textnormal{on } \gamma,\label{equ:ER2}\\
     (\vec{v}_\FF-\vec{v}_\PM^{\textnormal{int}} ) \cdot \vec{\tau}   &=  \epsilon N_1^\BL \frac{\partial \vec{v}_\FF}{\partial \vec{n}} \cdot\vec{\tau} \quad  &\textnormal{on } \gamma,\label{equ:ER3}
\end{align}
where the boundary layer coefficients $N_1^{\BL}>0$ and $N_\mathrm{s}^{\BL}$ are computed solving the appropriate boundary layer problems~\citep[Sec. 3.2.2]{Eggenweiler_Rybak_MMS20} for the given pore geometry. The interfacial porous-medium velocity in Eq.~\eqref{equ:ER3} is defined by 
\begin{equation}
    \vec{v}_\PM^{\textnormal{int}} := -\frac{\epsilon^2}{\mu} \sum_{j=1}^2 \ten{M}^{j,\BL} \frac{\partial p_\PM}{\partial x_j} \cdot \vec{\tau} \quad \textnormal{on } \gamma, \label{equ:ERinterfacialvelocity}
 \end{equation}
where $\ten{M}^{j,\BL} = \left(M_1^{j,\BL}, M_2^{j,\BL}\right)^\top$ for $j=1,2$ is the boundary layer constant defined in~\citep[Sec. 3.2.3]{Eggenweiler_Rybak_MMS20}
with $M_1^{j,\BL}>0$. 
\begin{remark}
    Note that the coupling condition~\eqref{equ:ER2} includes the correction term to the balance of normal forces in Eq.~\eqref{eqn:BJ2} for the case of anisotopic porous media. The interface condition~\eqref{equ:ER3} can be regarded as a jump of the tangential velocity component as the Beavers--Joseph condition~\eqref{eqn:BJcondition}.
\end{remark}

\section{Numerical results}\label{sec3:NumSimRes}
In this section, we study the considered coupling concepts numerically. We begin by 
validating the hybrid-dimensional model against an analytical solution.
Then, we present two test cases: lid--driven cavity  over porous bed and splitting flow system. 
For each test case, we validate the numerical results of the hybrid-dimensional model by comparing them to pore-scale-resolved simulations.
Additionally, we compare the proposed model with the Stokes--Darcy models, considering both classical and generalised interface conditions.

\begin{figure}[h!]
    \centering
    \includegraphics[width=0.8\linewidth]{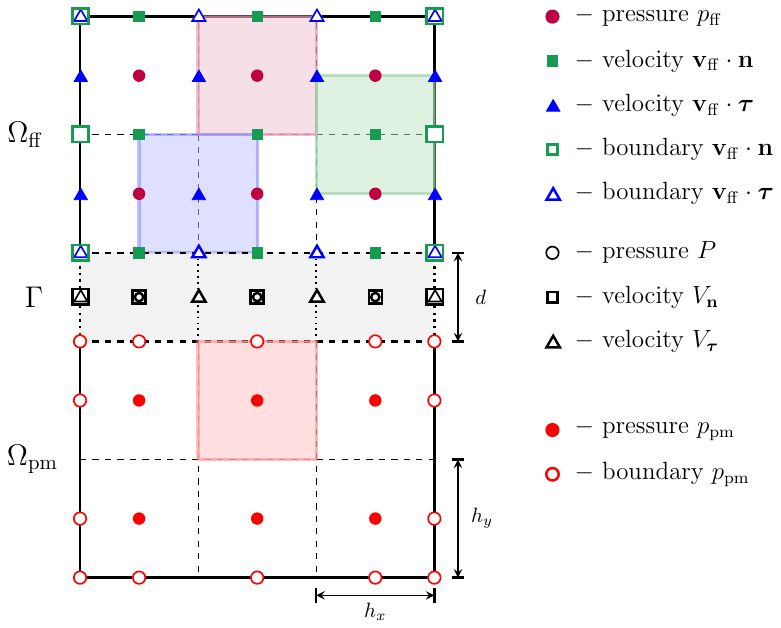}
    \caption{Staggered grid in the coupled domain for the hybrid-dimensional model}
    \label{fig:MACschme}
\end{figure}

The pore-scale problem \eqref{eqn:porescaleStokes1}--\eqref{eqn:porescaleBC} is solved using the software package FreeFEM++ with the Taylor--Hood finite elements \citep{Hecht_12}. The same software is applied to compute the permeability tensor and boundary layer constants appearing in the generalised interface conditions~\citep[Sec. 3.2]{Eggenweiler_Rybak_MMS20}.  The macroscale Stokes and Darcy problems \eqref{equ:Stokes1}--\eqref{equ:PMBC} are discretised by the second-order finite volume method on staggered rectangular grids conforming at the fluid--porous interface with the grid size $h_x$, $h_y$ (Fig.~\ref{fig:MACschme}). The problem is completed by the classical interface conditions \eqref{eqn:BJ1}--\eqref{eqn:BJcondition} or generalised conditions \eqref{equ:ER1}--\eqref{equ:ER3}. The proposed hybrid-dimensional model~\eqref{eqn:AverageBrinkmanMass}--\eqref{eqn:TransmissionGammaPM} is discretised using the second-order staggered finite difference scheme along the complex interface $\Gamma$ with the grid size $h_x$. The grids on the complex interface $\Gamma$ are conforming with the grids in the free-flow and porous-medium subdomains~(Fig.~\ref{fig:MACschme}).

\begin{remark}
Note that there is a gap of size $d$ between the free-flow and porous-medium regions due to the geometric configuration of the transition zone in the hybrid-dimensional model. We consider this gap as a shift either in the free-flow or porous-medium subdomain depending on the external boundary conditions. The thickness \mbox{$d>0$} is sufficiently small compared to the length of the flow domain, therefore the shift in the subdomain is negligible.
\end{remark}

\subsection{Validation against analytical solutions}
In our previous work~\citep{Ruan-Rybak-2024}, numerical solutions for the hybrid-dimensional model were validated against analytical solutions, demonstrating the second-order convergence of the discretisation scheme. In this section, we consider appropriate analytical solutions and study effects of the thickness $d$ and the profile of the normal velocity across the complex interface determined by $\lambda_1$, $\lambda_2$.

The geometry of the coupled flow domain is set as $\Omega = [0, 1] \times [0, 1+d]$, where the free-flow region $\Omega_\FF = [0, 1] \times [0.5+d,1+d]$ and porous medium $\Omega_\PM = [0,1] \times [0, 0.5]$ are separated by the complex interface $\Gamma = [0, 1 ] \times [0.5, 0.5+d]$. In this test case, the stress jump is disregarded  ($\ten{\beta} = \ten{0}$). The dynamic and effective viscosity is $\mu = \mu_\EF = 1$, and the slip coefficient is set to $\alpha=0.1$. The permeability tensors are $\ten{K}_\PM= \ten{K}_\Gamma= 10^{-2} \ten{I}$. 

The analytical solution for the hybrid-dimensional problem is constructed to satisfy the incompressibility condition in the free flow~\eqref{equ:Stokes1} and the transmission conditions~\eqref{eqn:TransmissionGammaFFNormal}--\eqref{eqn:TransmissionGammaPM}. The analytical solutions in the free-flow and porous-medium regions are given by
\begin{align}
     \vec{v}_\FF \cdot \vec{\tau}=\cos{ \left(x_1 \right) } \exp{( x_2-0.5)},\hspace{+4.4ex} \vec{v}_\FF\cdot \vec{n} =\sin{\left( x_1\right)} \exp{(x_2-0.5)},\hspace{+1.6ex} \nonumber\\
     p_\FF\, =\, \sin{( x_1 + x_2 - 0.5)},\hspace{+9.4ex} p_\PM\, =-100(x_2-0.5)\sin{( x_1)},\label{equ:exactsolution}
\end{align}
and on the complex interface $\gamma$:
\begin{align}
    V_\vec{\tau} = \cos{(x_1)}(\exp{(d)}-1)/d,\hspace{+2ex} V_\vec{n} =\sin{(x_1)}(\exp{(d)}-1)/d,\hspace{+6ex}\nonumber\\
    P = - (\cos{(x_1+d)}-\cos{(x_1)})/d. \label{equ:exactsolutiongamma}
\end{align} Substituting \eqref{equ:exactsolution}, \eqref{equ:exactsolutiongamma} into \eqref{equ:Stokes2}--\eqref{equ:Darcy1}, \eqref{equ:PMBC}, and \eqref{eqn:AveragedBrinkmanMomentumNormal}--\eqref{eqn:gammaBC}, the corresponding source terms and the boundary values are obtained. Here, Dirichlet boundary conditions are applied on the top of the free flow and the boundaries of the porous-medium subdomain, while Neumann boundary conditions are imposed on the boundary of the complex interface $\partial \Gamma$ and the remaining boundaries of the free-flow subdomain.

\begin{figure}[h!]
\centering
    \begin{subfigure}[b]{0.44\textwidth}
    \includegraphics[width=\textwidth]{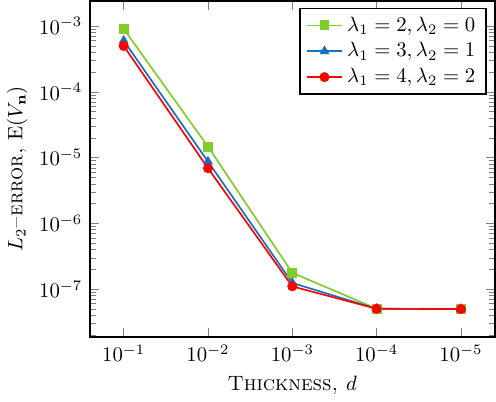}
    \caption{}
    \end{subfigure}
    \quad
    \begin{subfigure}[b]{0.44\textwidth}
    \includegraphics[width=\textwidth]{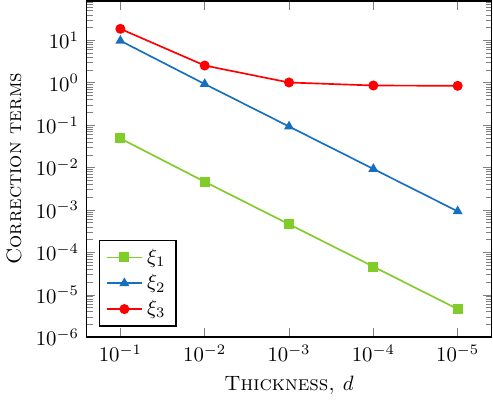}
        \caption{}
    \end{subfigure}
    \caption{$L_2$-error for averaged normal velocity for different $\lambda_1$, $\lambda_2$ (a) and correction terms $\xi_1$, $\xi_2$, $\xi_3$ for different thickness $d$ (b)}\label{fig:Anacpt}
\end{figure}

In section~\ref{sec:hybrid}, we introduced the dimensionless numbers $\lambda_1$ and $\lambda_2$, which determine the velocity profile in the normal direction. In this test case, we compare the numerical solution of the hybrid-dimensional model against the analytical solution for different normal velocity profiles. The $L_2$-error of the averaged normal velocity is $\mathrm{E}({V_\vec{n}}) = \|V_{\vec{n}}^h- V_{\vec{n}}\|_{L^2(\Gamma)} $, where $V_{\vec{n}}^{h}$ is the numerical solution. The grid size is $h_x=h_y=1/800$. Starting from $d=1/10$, the thickness of the complex interface is decreased by the factor of ten and five refinement levels are considered. The simulation results in Fig.~\ref{fig:Anacpt}a show that application of both piecewise linear ($\lambda_1=3$, $\lambda_2=1$) and quadratic ($\lambda_1=4$, $\lambda_2=2$) assumptions can describe normal velocity across the interface more accurately than the linear case ($\lambda_1=2$, $\lambda_2=0$). When the thickness of the complex interface $d$ is sufficiently small, all choices of normal velocity profile provide a good approximation to the analytical solution.

The correction terms in Eq.~\eqref{eqn:AverageBrinkmanMass}--\eqref{eqn:AveragedBrinkmanMomentumTangential} model storage and transfer of mass and momentum along $\Gamma$. We evaluate the effects of the correction terms by computing
\begin{equation}
    \xi_i := \int_\Gamma  |c_i|~\mathrm{d}x_1, \quad i \in \{1,2,3\},  
\end{equation}
where
\begin{align}
    c_1 &:=  -d \frac{\partial  V_{\vec{\tau}} }{\partial \vec{\tau}}, \quad c_2 := d \Big(  
    \mu (\ten{K}_{\Gamma}^{-1} \vec{V})\cdot \vec{n}- \mu_\EF \frac{\partial^2 V_{\vec{n}}}{\partial \vec{\tau}^2}-F_{\vec{n}}\Big),\\
    c_3 &:= d \Big( 
        \mu (\ten{K}_{\Gamma}^{-1} \vec{V}) \cdot  \vec{\tau}
        -  \mu_\EF \frac{\partial^2 V_{\vec{\tau}}}{\partial \vec{\tau}^2} + \frac{\partial P}{\partial \vec{\tau}}-F_{\vec{\tau}} \Big)+\frac{\alpha \mu_\EF (6 V_{\vec{\tau}}- 2 \vec{v}_\FF \cdot  \vec{\tau}|_{\gamma_\FF})}{\alpha d +4\sqrt{K_\PM}}.\label{eqn:cor3}
\end{align}

The simulation results in Fig.~\ref{fig:Anacpt}b indicate that the impact of the correction terms diminishes as the interface thickness decreases. Note that the last term in Eq.~\eqref{eqn:cor3} plays a dominant role when the complex interface becomes narrow.

\subsection{Lid-driven cavity over porous bed}\label{sec:lid-driven}
The lid-driven cavity flow is one of the most commonly used benchmarks in computational fluid dynamics due to its simple geometric setting and the intriguing flow patterns it produces.

\begin{figure}[h!]
    \centering
    \includegraphics[width = \textwidth]{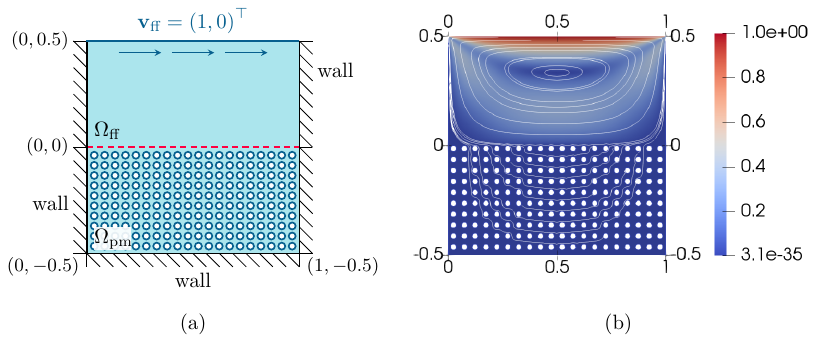}
    \caption{Lid-driven cavity: geometric setting (a) and pore-scale velocity magnitude (b)}\label{fig:LDC}
\end{figure}

In this test case, the size of the flow domain is set as $\Omega= [0, 1] \times [-0.5, 0.5]$ (Fig.~\ref{fig:LDC}a). On the macroscale, the free-flow and porous-medium subdomains are of equal size, with $\Omega_\FF = [0, 1] \times [0,0.5]$ and $\Omega_\PM = [0, 1] \times [-0.5,0]$ separated by a sharp interface $\gamma= (0,1)\times\{0\}$. In the hybrid-dimensional model, these subdomains are separated by a transition region $\Gamma = [0,1]\times[0,d]$, where $d=10^{-5}$. The presence of the transition zone results in a shift of the free-flow subdomain $\Omega'_\FF  = [0, 1] \times [d,0.5+d]$. 


In the pore-scale setting, the porous medium consists of $20 \times 10$ circular inclusions, with the pore size $\epsilon = 1/20$ (Fig.~\ref{fig:LDC}). The radius of each circular inclusion is set to $r = 0.25\epsilon$ leading to the porosity $\phi = (\epsilon^2 - \pi r^2) / \epsilon^2 = 0.8037$. According to~\eqref{eqn:unitcellpermeability}, the permeability tensor in the unit cell is determined to be  $\ten{K}=1.990\cdot 10^{-2}\ten{I}$. Consequently, the physical permeability tensor is $\ten{K}_\PM=\epsilon^2\ten{K}= 4.975\cdot 10^{-5} \ten{I}$. The fluid viscosity is set $\mu=1$. 

In the hybrid-dimensional model \eqref{eqn:AverageBrinkmanMass}--\eqref{eqn:TransmissionGammaPM}, we use the same permeability tensor in the transition region as in the porous medium, i.e., $\ten{K}_\TR=\ten{K}_\PM= 4.975\cdot 10^{-5} \ten{I}$. The effective viscosity is $\mu_\EF=\mu/\phi= 1.244$ and the slip coefficient is $\alpha = 1$. The stress jump tensor is set $\ten{\beta}= 4.0\, \ten{I}$. The value of parameter $\ten{\beta}$ is chosen by testing different flow scenarios and optimising the difference between the macroscale and averaged pore-scale results.  
For this test case, we choose the quadratic normal velocity profile ($\lambda_1 = 4$, $ \lambda_2 = 2$) which demonstrated the best accuracy (Fig.~\ref{fig:Anacpt}). For the classical interface conditions \eqref{eqn:BJ1}--\eqref{eqn:BJcondition}, we set the Beavers--Joseph parameter $\alpha_{\mathrm{BJ}}=1$, which is a common value used in the literature. For the generalised interface conditions~\eqref{equ:ER1}--\eqref{equ:ER3}, the corresponding boundary constants are calculated for the given pore geometry $N^\BL_\mathrm{s}=0$, $N_1^\BL=5.384\cdot10^{-2}$, $\ten{M}^{1,\BL}=(3.116\cdot10^{-3},0)^\top$, $\ten{M}^{2,\BL}=\vec{0}$. 
\begin{figure}[h!]
\centering
    \begin{subfigure}[b]{0.46\textwidth}
    \includegraphics[width=\textwidth]{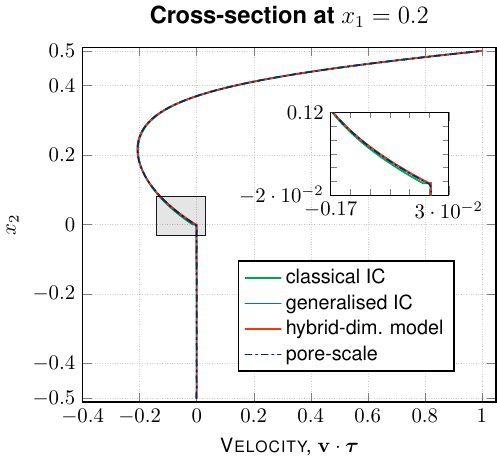}
        \caption{}
    \end{subfigure}
    \quad
    \begin{subfigure}[b]{0.46\textwidth}
    \includegraphics[width=\textwidth]{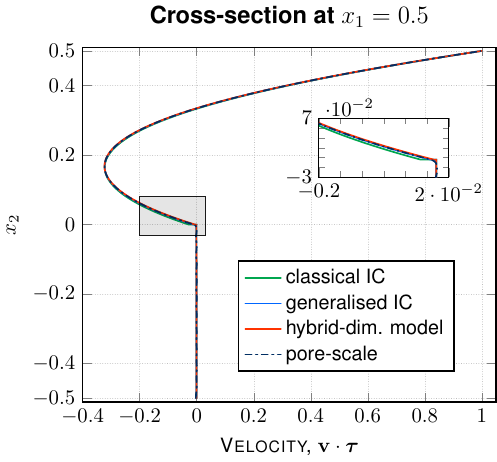}
        \caption{}
    \end{subfigure}
    \caption{Tangential velocity profiles for lid-driven cavity at $x_1 = 0.2$ (a) and $x_1=0.5$~(b)}\label{fig:LDCUprofile0205}
\end{figure}

For the lid-driven cavity, the boundary conditions are defined in Fig.~\ref{fig:LDC}a. On the top boundary, we consider $\vec{v}_\FF = \vec{v}_\epsilon = (1,0)^\top$. For the rest of the external boundaries, we have the ``wall" boundary conditions, which read in the pore-scale setting: 
\begin{align}
        \vec{v}_\epsilon = \vec{0} \quad \textnormal{on } \partial \Omega_\WALL, 
\end{align}
and in the macroscale setting: 
\begin{align}
    \vec{v}_\FF =\vec{0} \quad \textnormal{on } \Gamma_{\FF,\WALL},  \qquad \vec{v}_\PM  \cdot \vec{n}_\PM=0 \quad \textnormal{on } \Gamma_{\PM,\WALL}.
\end{align}
In the hybrid-dimensional model, the ``wall" condition is 
\begin{align}
    \vec{V} = \vec{0}\quad  \textnormal{on } \partial \Gamma_{\WALL}.
\end{align}
The pore-scale velocity magnitude and the streamline pattern of the lid-driven cavity are shown in Fig.~\ref{fig:LDC}b. The recirculating vortex is driven by the horizontal velocity on the top of the domain and the streamlines are nearly symmetric. The flow is mainly parallel to the fluid--porous interface in the centre of the domain and the normal velocity is small. The tangential velocity profiles for the pore- and macroscale models are compared at the cross-sections $x_1=0.2$ and $x_1 =0.5$ (Fig.~\ref{fig:LDCUprofile0205}). Here, the following notations are used: pore-scale resolved model (pore-scale), classical interface conditions (classical IC), hybrid-dimensional Stokes--Brinkman--Darcy model (hybrid-dim. model).  All the macroscale numerical simulation results show good agreement with the pore-scale results due to mainly parallel flow near the interface.

\subsection{Splitting flow}\label{sec:splittedFlow}
In this test case, we demonstrate that the hybrid-dimensional model is suitable for general flow problems, where the flows are arbitrary to the interface. We consider the same geometric setting as in section~\ref{sec:lid-driven}.

\begin{figure}[h!]
    \centering
    \includegraphics[width=\textwidth]{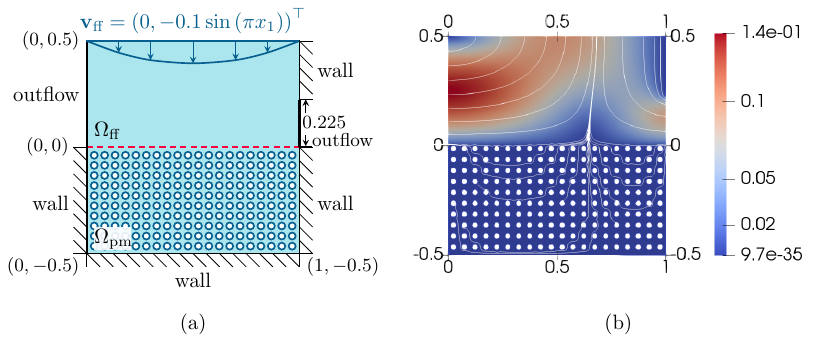}
    \caption{Splitting flow: geometric setting (a) and pore-scale velocity magnitude (b)}\label{fig:splittedflow}
\end{figure}

\begin{figure}[h!]
    \centering
    \begin{subfigure}[b]{0.45\textwidth}
    \includegraphics[width=\textwidth]{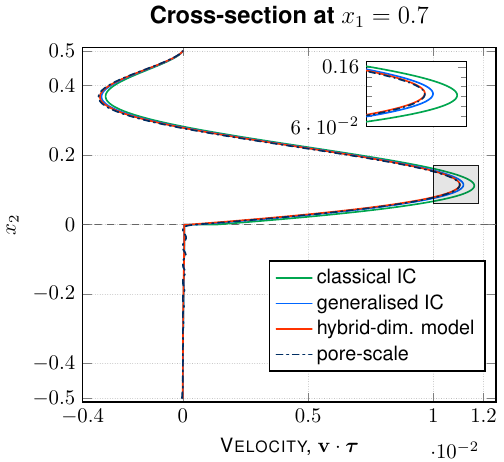}\label{fig:splittedflowUProfile07}
        \caption{}
    \end{subfigure}
    \quad
    \begin{subfigure}[b]{0.45\textwidth}
    \includegraphics[width=\textwidth]{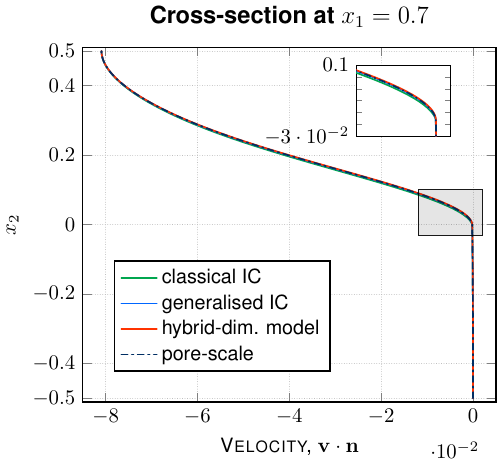}
        \caption{}
    \end{subfigure}
    \caption{Velocity profiles for splitting flow system at $x_1=0.7$}\label{fig:splittedflowvelocityprofile07}
\end{figure}

\begin{figure}[h!]
\centering
    \begin{subfigure}[b]{0.45\textwidth}
    \includegraphics[width=\textwidth]{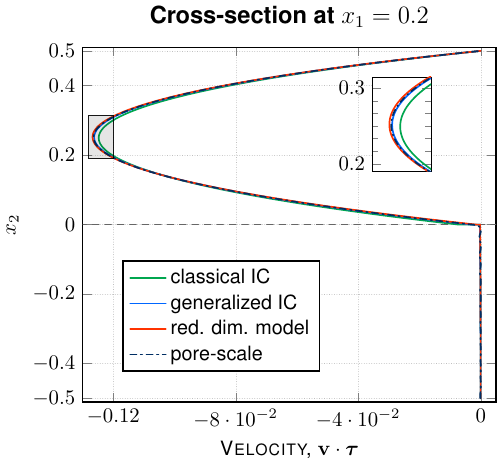}
        \caption{}
    \end{subfigure}
    \quad
    \begin{subfigure}[b]{0.46\textwidth}
    \includegraphics[width=\textwidth]{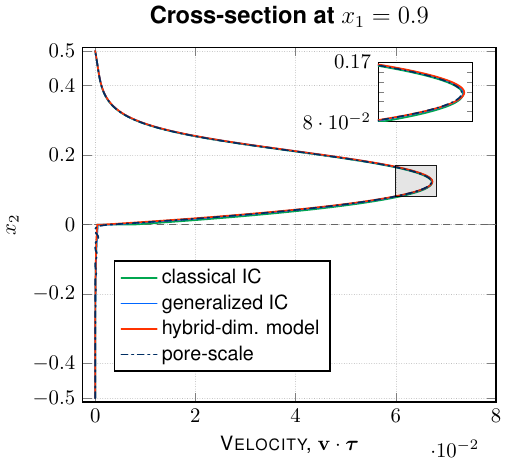}
        \caption{}
    \end{subfigure}
    \caption{Tangential velocity profiles at $x_1 = 0.2$ (a) and $x_1 = 0.9$ (b)}\label{fig:splittedflowvelocityprofile0209}
\end{figure}

The boundary conditions of this test case are schematically presented in Fig.~\ref{fig:splittedflow}a. The inflow velocity on the top of the free-flow subdomain is set as $\vec{v}_\FF = \vec{v}_\epsilon = (0, -0.1\sin{(\pi x_1)})^\top$. For the pore-scale model, we define the following ``outflow" boundary conditions
\begin{align}
  \mu\frac{\partial \vec{v}_\epsilon}{\partial \vec{n}}\cdot \vec{\tau} -p_\epsilon =0, \quad \vec{v}_\epsilon \cdot \vec{n} = 0 \quad \textnormal{on } \partial \Omega_\OUTFLOW.
\end{align}
In the macroscale setting, the ``outflow" boundary conditions are 
\begin{align}
     \mu \frac{\partial \vec{v}_\FF}{\partial \vec{n}}\cdot \vec{\tau} -p_\FF =0, \quad \vec{v}_\FF \cdot \vec{n} = 0  &\quad \textnormal{on }\Gamma_{\FF, \OUTFLOW},\\
    \mu_\EF\frac{\partial V_\vec{\tau}}{\partial \vec{n}} -P =0, \hspace*{+6ex} V_\vec{n} = 0  &\quad \textnormal{on }\Gamma_{\OUTFLOW}.
\end{align}
The pore-scale velocity field is visualised in Fig.~\ref{fig:splittedflow}b. We observe that in this case the flow is arbitrary to the fluid--porous interface and provide the profiles of both velocity components at different cross-sections (Fig.~\ref{fig:splittedflowvelocityprofile07}, Fig.~\ref{fig:splittedflowvelocityprofile0209}). 
Velocity profiles are plotted at $x_1 = 0.7$, where the flow splits near the interface. Note that for the normal component of velocity, all numerical results are close to the pore-scale results (Fig.~\ref{fig:splittedflowvelocityprofile07}b).  However, a significant difference can be observed for the tangential velocity at the cross-sections, where the flow is not nearly parallel to the interface. In Fig.~\ref{fig:splittedflowvelocityprofile07}a and Fig.~\ref{fig:splittedflowvelocityprofile0209}a, the results of the hybrid-dimensional model are almost identical to the pore-scale simulations. The generalised interface conditions show a good agreement with the pore-scale data as well. However, simulation results obtained using the classical interface conditions do not match so well with the pore-scale velocity profile.

We also provide the tangential velocity profiles for different models at $x_1 = 0.9$, where the flow is almost parallel to the fluid--porous interface (Fig~\ref{fig:splittedflowvelocityprofile0209}b). Here, the numerical simulation results for all macroscale models provide accurate results as in the case of the lid-driven cavity.  

According to \citet{Strohbeck-Eggenweiler-Rybak-23}, achieving accurate results for arbitrary flow directions requires different values of the Beavers--Joseph parameter at various positions along the interface. However, for the proposed hybrid-dimensional model, a single optimal choice of parameter $\ten{\beta}$ is sufficient to maintain accuracy across different cross-sections without compromising the fidelity of results when compared to pore-scale simulations. 




\section{Discussion and conclusions}\label{sec:discussion}
In this paper, we studied the hybrid-dimensional Stokes--Brinkman--Darcy model derived in our previous work by averaging the Brinkman equations across the transition region between the free flow and porous medium. This thin transition zone serves as a complex interface and involves higher-order correction terms that enhance accuracy and make the model suitable for arbitrary flow directions near the fluid--porous interface.

To validate the proposed model and to study the influence of the higher-order terms as well as different velocity profiles across the complex interface, we first choose appropriate analytical solutions. The model is more accurate by allowing non-constant velocity profiles in the complex interface.  The correction terms in the tangential direction play an essential role when the thickness of the complex interface decreases. Then, we consider two benchmarks and compare the developed hybrid-dimensional model and the Stokes--Darcy model with the classical and generalised coupling conditions against the pore-scale resolved model. These simulation results demonstrate the suitability of the hybrid-dimensional model for arbitrary flow directions to the interface in comparison to the classical conditions, and a similar accuracy (slightly more accurate) to the generalised interface conditions. These results demonstrate that the developed hybrid-dimensional model is an accurate alternative to the generalised interface conditions and it effectively handles arbitrary flow directions near the interface that is of highest importance for a wide range of applications. Future extension of this work will include the inertial terms in the hybrid-dimensional problem. 


\backmatter









\bibliography{FLUPOR}

\section*{Statements and Declarations}
\subsection*{Funding} The work is funded by the Deutsche Forschungsgemeinschaft (DFG, German Research Foundation) -- Project Number 490872182 and Project Number 327154368 -- SFB 1313.
Open Access funding is enabled and organized by Projekt DEAL. 

\subsection*{Competing Interests}
The authors have no relevant financial or non-financial interests to disclose.
\subsection*{Author Contributions} All authors contributed to the conceptualization and methodology. The derivation of the hybrid-dimensional model, the implementation and model validation were done by Linheng Ruan. The comparison study was performed by both authors. Linheng Ruan prepared the initial version of manuscript. Both authors finalized and approved the manuscript.

\subsection*{Data Availability} The datasets generated and analysed during the current study are available from the corresponding author on reasonable request.


\end{document}